\documentclass[fleqn,usenatbib]{mnras}

\usepackage{newtxtext,newtxmath}

\usepackage[T1]{fontenc}

\DeclareRobustCommand{\VAN}[3]{#2}
\let\VANthebibliography\thebibliography
\def\thebibliography{\DeclareRobustCommand{\VAN}[3]{##3}\VANthebibliography}

\def\cmsg{\, \mathrm{cm^2 \, g^{-1}}}

\usepackage{graphicx}	
\usepackage{amsmath}	

\title[Galaxies in SIDM]{Properties and observables of massive galaxies in self-interacting dark matter cosmologies}

\author[Claudio Mastromarino et al.]{Claudio Mastromarino,$^{1,2}$\thanks{E-mail: claudio.mastromarino96@gmail.com}
Giulia Despali,$^{3}$
Lauro Moscardini,$^{2,4,5}$
Andrew Robertson,$^{6}$ \newauthor
Massimo Meneghetti,$^{4,5}$
Matteo Maturi$^{3}$
\\
$^{1}$INFN-Sezione di Roma ‘Tor Vergata’, Via della Ricerca Scientifica, 1, 00133, Roma, Italy\\
$^{2}$Dipartimento di Fisica e Astronomia "A. Righi", Alma Mater Studiorum Universit\`a di Bologna, Via Piero Gobetti 93/2, I-40129 Bologna, Italy\\
$^{3}$Institut f\"{u}r Theoretische Astrophysik, Zentrum f\"{u}r Astronomie, Heidelberg Universit\"{a}t, Albert-Ueberle-Str. 2, 69120, Heidelberg, Germany\\
$^{4}$INAF-Osservatorio di Astrofisica e Scienza dello Spazio di Bologna, Via Piero Gobetti 93/3, I-40129 Bologna, Italy\\
$^{5}$INFN-Sezione di Bologna, Viale Berti Pichat 6/2, I-40127 Bologna, Italy\\
$^{6}$Jet Propulsion Laboratory, California Institute of Technology, 4800 Oak Grove Drive, Pasadena, CA 91109, USA\\
}

\date{Accepted XXX. Received YYY; in original form ZZZ}

\pubyear{2022}

\begin{document}
\label{firstpage}
\pagerange{\pageref{firstpage}--\pageref{lastpage}}
\maketitle

\begin{abstract}
We use hydrodynamical cosmological simulations to test the differences between cold and self-interacting dark matter models (CDM and SIDM) in the mass range of massive galaxies ($10^{12}\,M_{\odot}h^{-1}<M<10^{13.5}\,M_{\odot}h^{-1}$). We consider two SIDM models: one with constant cross section $\sigma/m_{\chi} = 1\cmsg$ and one where the cross section is velocity-dependent. We analyse the halo density profiles and concentrations, comparing the predictions of dark-matter-only and hydrodynamical simulations in all scenarios. We calculate the best-fit Einasto profiles and compare the resulting parameters with previous studies and define the best-fit concentration-mass relations. We find that the inclusion of baryons reduces the differences between different dark matter models with respect to the DM-only case. In SIDM hydro simulations, deviations from the CDM density profiles weakly depend on mass: the most massive systems ($M>10^{13}M_{\odot}h^{-1}$) show cored profiles, while the least massive ones ($M<10^{12.5}M_{\odot}h^{-1}$) have cuspier profiles.  Finally, we compare the predictions of our simulations to observational results, by looking at the dark matter fractions and the distribution of strong lensing Einstein radii. We find that in SIDM the dark matter fractions decrease more rapidly with increasing stellar mass than in CDM, leading to lower fractions at $M_{*}>10^{11}M_{\odot}$, a distinctive signature of self-interactions. At the same time, the distribution of Einstein radii, derived from both CDM and SIDM hydro runs, is comparable to observed samples of strong lenses with $M>10^{13}M_{\odot}h^{-1}$. We conclude that the interplay between self-interaction and baryons can greatly reduce the expected differences between CDM and SIDM models at this mass scale, and that techniques able to separate the dark and luminous mass in the inner regions of galaxies are needed to constrain self-interactions. 
\\



\end{abstract}

\begin{keywords}
cosmology: dark matter - methods: numerical - galaxies:halos -  galaxies: elliptical and lenticular, cD - X-rays: galaxies - gravitational lensing: strong
\end{keywords}



\section{Introduction}
The dominant hypothesis for the formation and evolution of cosmic structure in our Universe is the $\Lambda$ cold dark matter ($\Lambda$CDM) model, in which dark matter (DM) particles are non-relativistic and collisionless. This standard model is able to explain several fundamental properties of galaxy formation and evolution \citep{tesi:1,tesi:2} and it predicts the spectrum of matter fluctuations in the early Universe with exceptional accuracy \citep{tesi:Planck}. However, the absence of experimental evidence of collisionless cold dark matter particles \citep{tesi:WIMP}, and the fact that N-body simulations based on CDM models presented some discrepancies with observed quantities - such as the missing satellites \citep{tesi:MissSat}, core-cusp \citep{tesi:CoreCusp}, diversity \citep{tesi:Divesity} and Too-Big-To Fail \citep{tesi:TBTF} problems - generated interest in alternative dark matter models. The inconsistencies between simulations and observations have been in part mitigated with the inclusion of baryonic effects in simulations and with the discovery of new faint satellites in the Milky Way \citep{bechtol2015eight,drlica2015eight}. 
However, it is still uncertain if this is sufficient to bring the simulated CDM predictions completely in agreement with observations or if the discrepancies tell us something about the nature of dark matter, and thus it is the CDM model that needs to be revised. 

One possibility, first introduced by \citet{tesi:SIDM}, is that dark matter is not collisionless, but has self-interactions in addition to gravity. Dark matter self-interactions can have an impact on the macroscopic characteristics of haloes, alleviating some of the issues that emerge in a collisionless CDM scenario, while leaving the properties on large scale unchanged; in particular, self-interactions can flatten the centrally peaked cusps in the inner regions of galaxies and disturb the growth of dense satellite galaxies predicted by simulations, that are at the basis of the core-cusp and too-big-to-fail issues \citep{tesi:Solve1,tesi:Solve2}. Recently, self-interacting dark matter (SIDM) has received new attention thanks to the numerical developments that now allow us to accurately simulate these scenarios, modelling elastic or inelastic particle scattering, constant or velocity-dependent self-interaction cross section or more exotic variations \citep{tesi:Solve1,vogel16,vogel19,tesi:Elliptical5,tesi:Elliptical2,sameie20,rocha13,tesi:Robertson1,robertson21,lovell19,kaplinghat14,kaplinghat19,kaplinghat20}.


Constraints on the self-interaction cross-section on different scales have often been set by comparing observations to simulated predictions that do not include the physics of baryons. In these scenarios the distribution of DM in dwarf galaxies requires a cross-section $\sigma/m > 0.5 \cmsg$ \citep{tesi:GalassieNane1}, while from the ellipticity of elliptical galaxies it was deduced that $\sigma/m \leq 1\cmsg$ \citep{tesi:Elliptical7} and strong lensing arc statistics in galaxy clusters allow $\sigma/m < 0.1 \cmsg$ \citep{tesi:Cluster3}. Given these disparities on different scales, a constant cross-section compatible with cluster scale limitations is unable to appreciably lower the central density of dwarf galaxies \citep{tesi:vdcross1}. As a result, there has been a surge in interest in SIDM models with velocity-dependent cross-sections, i.e a cross-section that decreases with increasing relative velocity of the particles. The effective cross section in dwarf galaxies can then be several orders of magnitude bigger than in cluster-sized halos, bringing constraints on different scales into agreement \citep{correa21}. However, the majority of this prior research has relied on DM-only simulations. Recent works are instead starting to model the interplay between self-interactions and baryonic physics, finding that previous predictions based on DM-only simulations need to be updated. \citet{tesi:Elliptical2} and \citet{tesi:Elliptical8} used simulations of isolated galaxies to show that the cross talk between SIDM and baryons produces a wide range of halo profiles, depending on how centrally concentrated the baryonic component is. \citet{tesi:Elliptical5} ran full-hydrodynamic zoom-in simulations of nine haloes hosting massive galaxies, finding that SIDM haloes can be be both cored and cuspy, dependent on halo mass, morphological type, as well as the halo mass accretion history; moreover, the same simulations show that CDM and SIDM halo shapes are similar when baryons are included \citep{despali22} and thus $\sigma/m = 1\cmsg$ is not yet excluded, revising the constraints from \citet{tesi:Elliptical7}. \citet{tesi:Robertson1, robertson18b,robertson21,shen22} have studied the effect of SIDM on the density profiles of galaxy clusters, also concluding that the inclusion of baryons reduces the difference between CDM and SIDM predictions. These results show that baryons can have a strong impact on the distribution of SIDM, similarly to what has been demonstrated in CDM simulations.

Moreover, \citet{tesi:Robertson1} demostrated that using only a small sample of simulated halos can lead to misleading results. For this reason, here  we select haloes hosting massive galaxies ($10^{12}M_{\odot}h^{-1}<M_{200}<10^{13.5}M_{\odot}h^{-1}$) in a cosmological box from \citet{robertson21} to study the interplay between baryonic physics and SIDM and test the results by \citet{tesi:Elliptical5} on a larger sample. We analyse the properties of massive galaxies in three scenarios: the standard CDM and two SIDM models, with a constant cross-section $\sigma/m=1 \cmsg$ or a velocity-dependent cross-section. 
We study the structure of haloes, in terms of their total and dark matter density profiles, we calculate the best-fit Einasto parameters and discuss the resulting concentration-mass relations. Finally, we discuss observable properties that could be used to distinguish CDM from the alternative models considered here, i.e. dark matter fractions and the sizes of the strong lensing Einstein radii - a proxy for a system's lensing power. 

The paper is structured as follows: in Section \ref{sec:sim} we describe the simulations and the halo selection, in Section \ref{sec:profiles} we present our results on the halo density profiles and concentration. In Section \ref{sec:fractions} we then discuss how the dark matter fractions depend on the dark matter model and how they compare to observed measurements, while in Section \ref{sec:lensing} we discuss the effect of different dark matter models on the lensing signal. Finally, we summarise our results and draw our final conclusions in Section \ref{sec:conclusions}.

\section{Simulations} \label{sec:sim}

The simulations used in this work were first presented in \citet{robertson21} and are part of the extended family of EAGLE simulations \citep{tesi:EagleSImulation}.
These are cosmological hydrodynamical simulations of galaxy formation run with the GADGET-3 code and a galaxy-formation model that includes gas cooling, star formation, and feedback from both stars and active galactic nuclei. 

Here we use a set of six runs in total, which simulate the same volume of V = (50cMpc)$^3 $ (L050N0752) in CDM and two SIDM models: one with constant cross section $\sigma/m = 1\,\cmsg$ and one with a velocity-dependent cross section. The implementation of DM self-interactions used in the SIDM simulations is presented in \citet{2017MNRAS.465..569R} and its integration into the EAGLE simulations discussed in \citet{tesi:Robertson1}. The velocity-dependent cross-section is described by 3 parameters: the dark matter mass $m_{\chi}$, the
mediator mass $m_{\phi}$ , and a coupling strength $\alpha_{\chi}$. The model corresponds to dark matter particles scattering though a Yukawa potential and the differential cross-section is 
\begin{equation}
\frac{d\sigma}{d\Omega}	= \frac{\alpha_{\chi}^2}{m_{\chi}^2(m_{\phi}^2/m_{\chi}^2 + v^2 \sin \frac{\theta}{2})^2}\,,
\end{equation}
where $v$ is the relative velocity between two dark matter particles, and $\theta$ is the polar scattering angle.
By defining $\omega = m_{\phi} c/ m_{\chi}$ as a characteristic velocity below
which the scattering is roughly isotropic with $\sigma \sim \sigma_0$, and above which the cross-section decreases with increasing velocity, the cross-section can be written as
\begin{equation}
    \frac{d\sigma}{d\Omega}	= \frac{\sigma_0}{4\pi(1 +\frac{v^2}{\omega^2}\sin \frac{\theta}{2})^2}\,.
\end{equation}
The velocity-dependent model considered here, already used in \citet{tesi:Robertson1}, has $m_{\chi} = 0.15\,GeV$, $m_{\phi}= 0.28 \,keV$ and $\alpha_{\chi} = 6.74 \times 10^{-6}$, coresponding to $\sigma_0 = 3.04\cmsg$, $\omega = 560\, km/s$.

Both a DM-only and a full-hydro version of the same box were run for all dark matter models, allowing us to study the effect of baryons and alternative dark matter at the same time. The simulations follow $N = 752^3 $ dark matter particles and an (initially) equal number of gas particles. The dark matter mass resolution is $m_{\text{DM}}= 9.70\cdot10^6 M_{\odot}$, while the baryonic resolution is $m_g = 1.81\cdot10^6 M_{\odot}$. The spacial resolution, i.e. the Plummer-equivalent gravitational softening length is $\epsilon = 0.7 \, \mathrm{kpc}$. The cosmological parameters from the Planck survey \citep{ade2014planck} are used: $\Omega_{m}=0.307$, $\Omega_{\Lambda}=0.693$, $\Omega_{b}=0.0483$, $\sigma_8 =0.83$ and $h=0.6777$.

All simulations were run with the fiducial EAGLE galaxy formation model \citep{tesi:EagleSImulation}. Throughout the paper, we label the full-hydrodynamic simulations run with a 'b' to distinguish them from their DM-only counterparts: CDM and CDMb, SIDM1 and SIDM1b, vdSIDM and vdSIDMb. These simulations and dark matter models have already been analysed in \citet{robertson21} to look at the density profiles of clusters of galaxies in CDM and SIDM models and in \citet{tesi:Robertson2} to analyze how SIDM effects behave over a wide range of mass scales.

\subsection{Halo selection}\label{Halo_Selection}

\begin{figure}
    \centering
    {\includegraphics[width=\linewidth]{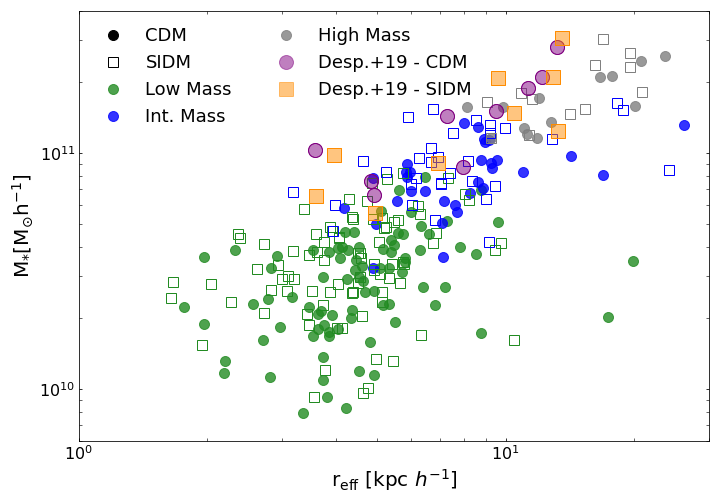}  }  
    \caption{Total stellar mass as a function of the effective radius of the central galaxy for the considered sample of haloes in the hydro runs at $z=0$. We compare them with the galaxies simulated (at $z=0.2$) by \citet{tesi:Elliptical5}. The distribution at the other two considered redshifts ($z=0.5,1$) is similar. The effective radius is calculated as the radius enclosing half of the stellar mass.}
    \label{fig:galaxies}
\end{figure}
\begin{table}
	\caption{Number of halos in the three bins in mass (Low Mass, Intermediate Mass and High Mass bin) in each of the six considered boxes (both DM-only and full-hydrodynamic simulations).}
\begin{tabular}{c @{\hspace{0.6\tabcolsep}}c@{\hspace{1.0\tabcolsep}}c@{\hspace{1.1\tabcolsep}}c@{\hspace{1.1\tabcolsep}}c@{\hspace{1.1\tabcolsep}}c}
	    \hline
	    &&\textbf{Low Mass} &  \textbf{Int. Mass} &\textbf{High Mass}&\textbf{Tot} \\
        &$\log(\frac{M_{200}}{M_{\odot}h^{-1}})$& [12,12.5] & [12.5,13] & [13,13.5]& [12,13.5]\\ \\
        \hline
        &\textbf{CDM} & 99 & 37 & 17 & 153  \\
		&\textbf{CDMb} & 83 & 32 & 14 &129\\
		$z=0$&\textbf{SIDM1} & 99 & 38 & 17&154 \\
		&\textbf{SIDM1b}& 83& 32& 13& 128\\
		&\textbf{vdSIDM}& 94 & 38 & 17 & 149\\
		&\textbf{vdSIDMb} & 87& 31& 15& 133\\
	    \hline
	    \hline
        &\textbf{CDM} &112 & 35 & 11 & 158  \\
		&\textbf{CDMb}& 94 & 31 &10 &135 \\
		$z=0.5$&\textbf{SIDM1}& 112& 35 & 11 & 158 \\
		&\textbf{SIDM1b}& 96& 31 & 10 & 137\\
		&\textbf{vdSIDM}& 113 & 34 & 11 & 158 \\
		&\textbf{vdSIDMb}& 94 & 32 & 10 & 136\\
		\hline
		\hline
        &\textbf{CDM} & 94 & 31 & 5 & 130   \\
		&\textbf{CDMb}& 80 & 27 & 4 & 111\\
		$z=1$&\textbf{SIDM1}&95 & 31 & 5 & 131 \\
		&\textbf{SIDM1b}& 80 & 28 & 4 & 112\\
		&\textbf{vdSIDM}& 97 & 31 & 5 & 133 \\
		&\textbf{vdSIDMb}& 82 & 28 & 4 & 114\\
		\hline
\end{tabular}	
	\label{tab:3}
\end{table}
The haloes are identified by SUBFIND, following the Friends-of-Friends (FOF) algorithm, with a linking length of $b = 0.2$ \citep{tesi:FOF}. 
We adopt $M_{200}$ as our definition of halo mass (i.e. the corresponding radius $R_{200}$ encloses 200 times the critical density $\rho_{c}(z)$) and we select systems with masses $10^{12}\,M_{\odot} h^{-1}< M_{200} <10^{13.5} \, M_{\odot} h^{-1}$. Throughout the paper, we split our sample into three mass bins: 
$12.0 < \log(M_{200}/M_{\odot}h^{-1}) < 12.5$ (Low Mass), $12.5 < \log(M_{200}/M_{\odot}h^{-1})< 13.0$ (Intermediate Mass) and $13.0 < \log(M_{200}/M_{\odot}h^{-1}) < 13.5$ (High Mass - up to the largest mass found in the box). The number of selected haloes in each mass bin and simulation are listed in Table ~\ref{tab:3}. 

We choose the mass range of massive galaxies to compare with previous predictions \citep{tesi:Elliptical5} and observational results \citep{auger10b,barnabe11,cappellari13,lovell18,sonnenfeld13} at the scales of elliptical galaxies. At redshift $z>0$, elliptical galaxies are the typical strong lenses: gravitational lensing is the most accurate technique to measure the total mass distribution of galaxies and clusters and thus it is especially relevant in the study of alternative dark matter models. We thus study both the properties of local galaxies at $z=0$ and of their counterparts at $z=0.5$ and $z=1$. Figure \ref{fig:galaxies} shows the galaxy properties of the considered sample at $z=0$ in the $M_{*}-r_{\text{eff}}$ plane, together with the sample simulated in \citet{tesi:Elliptical5}, demonstrating the extended statistics of this work. 

\begin{figure}
    \centering
    {\includegraphics[width=\linewidth]{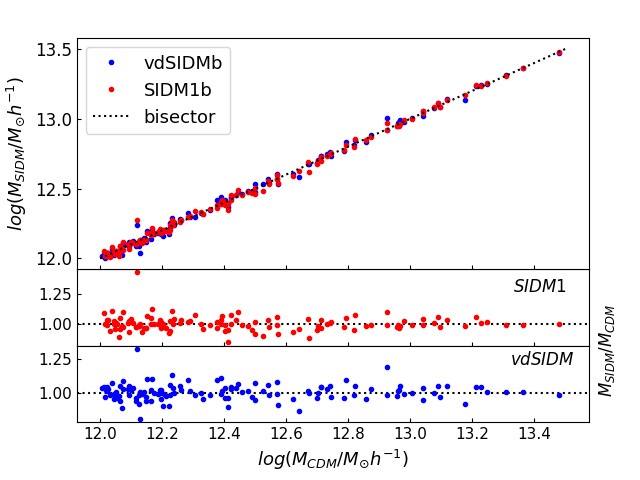}  }  
    \caption{Top: Masses of matched systems in full-hydrodynamic simulations at $z=0$; in blue matched system between CDMb and vdSIDMb,
    in red matched systems between CDMb and SIDM1b. The black dotted line represents the
    bisector.
    Middle: $M_{\text{SIDM1b}}/M_{\text{CDMb}}$. Bottom: $M_{\text{vdSIDMb}} /M_{\text{CDMb}}$ .}
    \label{fig:matched}
\end{figure}

We study the summary properties of haloes, but we also compare individual system to their counterparts across models. In order to identify the same objects in the six boxes, we match the position of the center of mass of each halo. We check that the distance between the matched haloes is smaller than their virial radius, i.e. $d_{\text{min}}/r_{\text{vir}} < 1$. \\
Figure \ref{fig:matched} shows the masses of matched haloes at $z=0$. Most of the haloes are distributed on the bisector, showing that corresponding systems have comparable masses and that matches have been correctly identified.


\section{Halo profiles and concentrations}\label{sec:profiles}
The influence of self-interactions on dark matter density profiles has been the driving force behind the field: its primary signature is a decrease in the dark matter density in the central regions of a halo when compared to CDM, due to the formation of a core. However, previous works have demonstrated that this effect can be mitigated or even reversed in the presence of a dense baryonic component \citep{tesi:SIDM+baryons,tesi:Elliptical5}. In this Section, we measure the density profiles and concentrations of the haloes in all runs and discuss the impact of SIDM and baryons.

\subsection{Method}
We calculate the spherically-averaged total and dark matter density profiles of individual haloes in each of the six boxes (see Section \ref{sec:sim}). We compute the density ($\rho_i$) in 30 logarithmically-spaced spherical shells. We use the radial range from $r=1 \,\mathrm{kpc\,h^{-1}}$ to the mean $R_{200}$ of each mass bin. The softening length of the simulations ($\epsilon = 0.7 \, \mathrm{kpc}$) is thus smaller than the minimum considered radius. 
We calculate the average profile for each mass bin and scenario, and fit it with an Einasto profile \citep{tesi:Einasto}, defined as
\begin{equation}
\label{Einasto}
\rho(r) = \rho_{-2}\,\, \exp \left\{-\frac{2}{\alpha_{\text{E}}}\left[\left(\frac{r}{r_{-2}}\right)^{\alpha_{\text{E}}}-1\right]\right\},
\end{equation}
where $r_{-2}$ and $\rho_{-2}$ are respectively the radius and the density at which $\rho(r)\propto r^{-2}$, while $\alpha_{\text{E}}$ is the shape parameter. The Einasto profile can reproduce a variety of shapes and it is thus a suitable match for both CDM and SIDM haloes, while the standard NFW profile \citep{tesi:NFW} cannot reproduce cored profiles. We also fit the individual haloes in each scenario and then use the best fit parameters to calculate the concentration $c_{200} = r_{200}/r_{-2}$.

\begin{figure}
    \centering
    {\includegraphics[width=\linewidth]{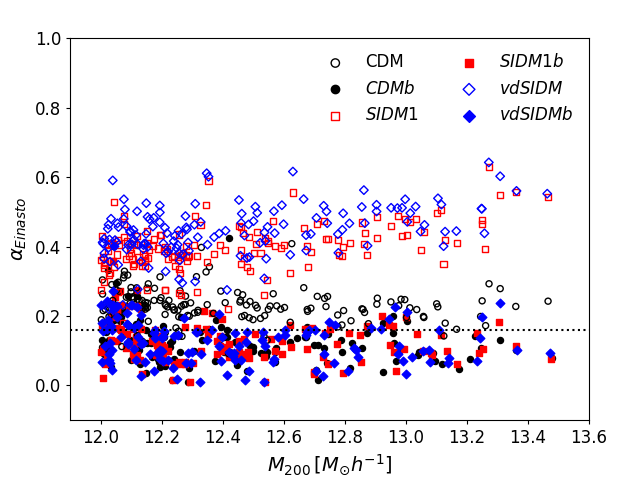}  }  
    \caption{The Einasto shape parameter $\alpha_{\text{Einasto}}$ for individual simulated haloes from different simulations. The empty points represent the best-fit values for DM-only runs with each dark matter model, while the filled points are the full-hydrodynamic simulations. The horizontal dotted line is fixed at $\alpha = 0.16$.}
    \label{fig:alpha}
\end{figure}

We perform the Einasto fit twice, both leaving the shape parameter free to vary and fixing it to $\alpha_{\text{E}}=0.16$, as proposed by previous works \citep{springel08b}. The Einasto best fit parameters, found as the values that minimize the sum of the squared residuals of $\log(\rho_{\text{model}})-\log(\rho_i)$, and related uncertainties for both cases can be found in Table ~\ref{tab:4}. The Einasto profile with  $\alpha_{\text{E}}=0.16$ is a good fit to CDM haloes and all full-hydro runs, but it fails to reproduce the central core the dark-matter-only SIDM runs, generating excedingly large scale radii. In this scenario, the fit is more precise when $\alpha_{\text{E}}$ is free to vary.
Fig.~\ref{fig:alpha} shows the distribution of the best-fit values of $\alpha$ as a function of mass for haloes at $z=0$. 
In the DM-only runs,  $\alpha_{\text{Einasto}}$ is higher in SIDM1 and vdSIDM (red and blue empty squares) than in CDM (black empty circles), due to the presence of the central core. Instead, when baryons are included the best-fit values are similar in all models (black, blue and red filled symbols) and closer to $\alpha=0.16$. In the rest of the paper we show results from the free-$\alpha$ fit, given that performs equally well on all runs.

\begin{table*}
	\centering
	\caption{Best fit parameters with Einasto profile for the three mass bins at $z=0$, for each simulation.  $\rho_{-2}$ is expressed in $M_{\odot}h^{-1}/(kpc/h)^3$, $r_{-2}$ in $kpc/h$ and $\alpha$ is dimensionless. The errors represent 1 standard deviation. On the left the fit is made leaving $\alpha$ as a free parameter, while on the right $\alpha$ is fixed to $0.16$.}
	\begin{tabular}{cccccccc}
	    \hline
	    &&&\textbf{Free $\alpha$} &&& \textbf{Fixed $\alpha$}\\
		\hline
		& &  \textbf{Low Mass} &  \textbf{Int. Mass} & \textbf{High Mass}&  \textbf{Low Mass} &  \textbf{Int. Mass} & \textbf{High Mass}\\
		\hline
		\textbf{CDM} & $\rho_{-2}$ &$(4.34 \pm 0.66)\cdot 10^6$ & $(3.90 \pm 0.72)\cdot 10^6$ & $(2.32 \pm 0.34)\cdot 10^6$  &$(4.55 \pm 1.37)\cdot 10^6$ & $(3.29 \pm 1.12)\cdot 10^6$ & $(1.47 \pm 0.30)\cdot 10^6$\\
		& $r_{-2}$ &$19.96\pm1.51$ & $30.73\pm 2.68$ & $51.68\pm3.23$  &$16.95\pm2.39$ & $27.76\pm 4.89$ & $58.46\pm6.37$ \\
		& $\alpha$ & $0.26\pm0.02$ & $0.28\pm0.03$ & $0.23\pm0.02$ &&& \\
		\hline
		\textbf{CDMb} & $\rho_{-2}$ & $(1.31 \pm 0.39)\cdot 10^7$ & $(7.44 \pm 2.36)\cdot 10^6$ & $(3.60 \pm 1.06)\cdot 10^6$ & $(1.22 \pm 0.28)\cdot 10^7$ & $(7.22 \pm 2.02)\cdot 10^6$ & $(4.31 \pm 0.94)\cdot 10^6$\\
		& $r_{-2}$& $ 11.18 \pm1.75$ & $20.51\pm3.32$ & $38.01\pm5.29$& $ 11.78 \pm1.20$ & $21.36\pm2.76$ & $37.18\pm4.00$ \\
		& $\alpha$  & $0.15\pm0.02$ & $0.14\pm0.03$ & $0.12\pm0.02$  \\
		\hline
		\textbf{SIDM1}  & $\rho_{-2}$&$(2.42 \pm 0.33)\cdot 10^6$ & $(1.87 \pm 0.27)\cdot 10^6$ & $(1.21 \pm 0.14)\cdot 10^6$ &$(1.38 \pm 0.69)\cdot 10^6$ & $(4.87 \pm 2.81)\cdot 10^5$ & $(9.61 \pm 4.79)\cdot 10^4$\\
		& $r_{-2}$&$28.13\pm1.75$ & $47.85\pm 2.80$ & $ 81.58\pm3.52$ &$27.71\pm6.98$ & $63.45\pm 20.18$ & $ 223.27\pm73.19$\\
		& $\alpha$  &$ 0.39\pm0.03 $ & $0.46\pm0.04$ & $0.47\pm0.03$  \\
		\hline
		\textbf{SIDM1b} & $\rho_{-2}$ &$(1.89 \pm 0.67)\cdot 10^7$& $(7.35 \pm 2.18)\cdot 10^6$& $(2.41 \pm 0.70)\cdot 10^6$&$(1.58 \pm 0.38)\cdot 10^7$& $(7.19 \pm 1.93)\cdot 10^6$& $(2.93 \pm 0.60)\cdot 10^6$\\
		& $r_{-2}$&$9.56\pm1.75$ & $20.97\pm3.18$ &$46.34\pm6.21$ &$10.71\pm1.10$ & $21.63\pm2.69$ &$44.10\pm4.58$\\
		& $\alpha$ & $0.14\pm0.02$ & $0.15\pm0.02$ & $0.13\pm0.02$  \\
		\hline
		\textbf{vdSIDM} & $\rho_{-2}$ &$(2.04 \pm 0.28)\cdot 10^6$ & $(1.61 \pm 0.22)\cdot 10^6$ & $(1.12 \pm 0.14)\cdot 10^6$ &$(8.50 \pm 4.75)\cdot 10^5$ & $(2.95 \pm 1.84)\cdot 10^5$ & $(6.64 \pm 3.53)\cdot 10^4$ \\
		& $r_{-2}$ &$31.09\pm 1.84$ & $52.13\pm2.88$ & $85.70\pm3.77$  &$34.27\pm 10.03$ & $80.29\pm28.68$ & $273.26\pm98.91$ \\
		& $\alpha$ & $0.43\pm0.04$ & $0.50\pm0.04$ & $0.50\pm0.03$  \\
		\hline
		\textbf{vdSIDMb} & $\rho_{-2}$ & $(1.91 \pm 0.67)\cdot 10^7$ & $(8.48 \pm 2.84)\cdot 10^6$ & $(2.37 \pm 0.67)\cdot 10^6$ & $(1.56 \pm 0.36)\cdot 10^7$ & $(8.51 \pm 2.73)\cdot 10^6$ & $(2.77 \pm 0.57)\cdot 10^6$\\
		& $r_{-2}$& $ 9.42\pm1.72 $ & $19.66\pm3.4$ & $45.98\pm6.00$ & $ 10.70\pm1.08 $ & $19.59\pm2.87$ & $44.18\pm4.62$\\
		& $\alpha$  & $0.14\pm0.02$ & $0.16\pm0.03$ & $0.13\pm0.02$  \\
		\hline
		
	\end{tabular}

	\label{tab:4}\centering
\end{table*}

\subsection{Density profiles} \label{sec:profiles0}

\begin{figure*}
    \centering
    \includegraphics[width=1.0\linewidth,height=0.6\linewidth]{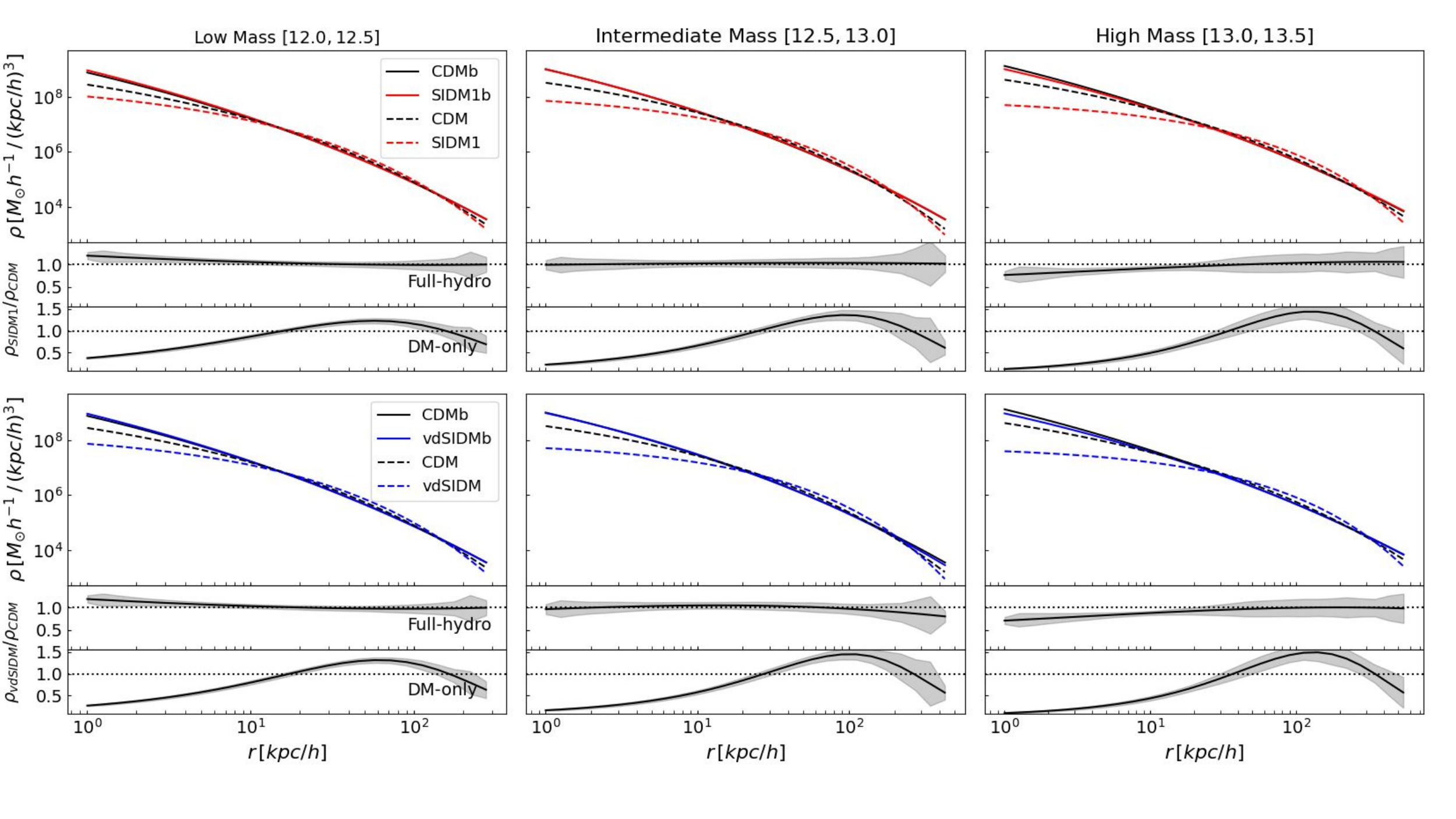}
    	\caption{Top: Fit of the total density profiles with CDM and SIDM1 models for both DM-only (dashed lines) and full-hydrodynamic (solid lines) simulations, at $z = 0$. The fit is performed on the mean profile of each bin. 
    	Bottom: ratio between the total and the DM-only profiles in the two models: $\rho_{\text{SIDM}}/\rho_{\text{CDM}}$. The shaded regions show the $1 \sigma$ uncertainties calculated on each mass bin and propagated on the ratios between the stacked profiles. The first density value is calculated at $r=1$ kpc $h^{-1}$, therefore above the spatial resolution (softening length $\epsilon = 0.7 \, \mathrm{kpc}$), for this reason this scale is not represented in these plots. The mass bins are in order, Low Mass: $\log(M_{200}/M_{\odot}h^{-1}) \in [12.0,12.5]$, Int. Mass: $\log(M_{200}/M_{\odot}h^{-1}) \in [12.5,13.0]$ and High Mass: $\log(M_{200}/M_{\odot}h^{-1}) \in [13.0,13.5]$.}
	\label{fig:densityz0}
\end{figure*}

\begin{figure*}
    \includegraphics[width=1.0\linewidth,height=0.35\linewidth]{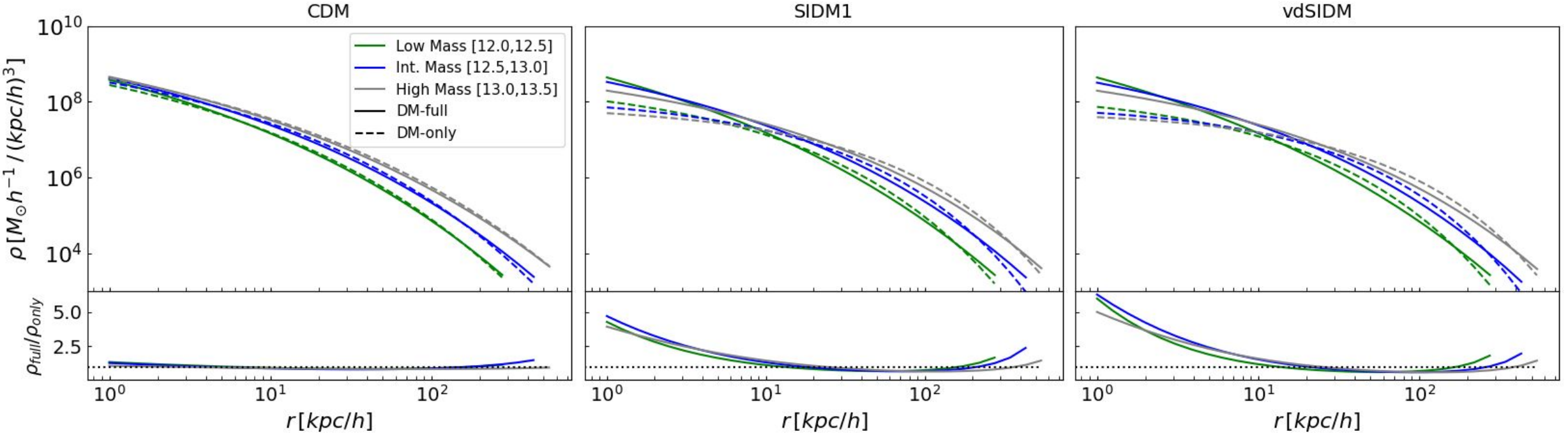}
	
	\caption{Top: dark matter density profiles for the three mass bins both for DM-only (dashed lines) and full-hydrodynamic (solid lines) simulations. The fit is performed on the mean profile of each bin. Bottom: ratio between dark matter density in full-hydrodynamic simulations $\rho_{\text full}$ and density in DM-only simulations $\rho_{\text only}$. On the right in the CDM model, in the middle  the SIDM1 model and on the left the vdSIMD model. For all models the addition of baryons modifies the distribution of dark matter by increasing its central density, but this effect is stronger for SIDM models ($\rho_{\text full}/\rho_{\text only} > 4.0$), than in CDM ($\rho_{\text full}/\rho_{\text only} < 1.5$).}
	\label{fig:Profiles_DM}
\end{figure*}

We start from the haloes at $z=0$ and we calculate the mean best-fitting Einasto profile for each considered mass bin. 
The results are shown in Fig.~\ref{fig:densityz0} for all models: CDM in black, SIDM1 in red and vdSIDM in blue. Here we show the total density profile, including both the baryonic and dark matter components in the full-hydro runs. Different columns show the results for the three mass bins, with solid (dashed) lines standing for full-hydro (DM-only) runs. In the middle (bottom) residual panels we plot the ratio between the SIDM and CDM profiles in the hydro (DM-only) runs. Finally, the 1$\sigma$ uncertainty on the ratio between the stacked profiles is represented by the gray shaded region.

In DM-only simulations, we recover the well-known trend of SIDM1 (and vdSIDM) haloes forming a central core: the central density ($r \leq 10 \, $kpc/h)  is lower by 50\% or more compared to the CDM case. In the full-hydro runs, the presence of baryons counteracts the core formation. Given that this effect is stronger for SIDM, baryonic physics almost erases the differences between CDM and SIDM profiles (see the residual panels in Figure \ref{fig:densityz0}). Our results go in the same direction as previous studies \citep{tesi:Elliptical2, tesi:Elliptical5, robertson21} 
and can be understood by considering that $M\sim10^{13}M_{\odot}h^{-1}$ corresponds to the halo mass in which star formation is most efficient, baryons dominate the inner density profile and the nature of dark matter plays a secondary role.

Despite this, we find a weak trend with mass:
\begin{itemize}
	\item \textbf{Low Mass} $(\log[M_{200}/M_{\odot}h^{-1}] \in [12.0,12.5])$: the SIDM1b and vdSIDMb profiles have a \emph{higher} density than that of CDMb ($\rho_{\text{ SIDM}}/\rho_{\text{CDM}} > 1$) in the central regions. However, the difference between the two models drops to less than 20\%.
	\item  \textbf{Intermediate Mass} $(\log[M_{200}/M_{\odot}h^{-1}] \in [12.5,13.0])$ : the difference between the  density profiles disappear and the profiles are almost exactly overlapping  ($\rho_{\text{ SIDM}}/\rho_{\text{CDM}} \sim 1$). 
	\item \textbf{High Mass} $(\log[M_{200}/M_{\odot}h^{-1}] \in [13.0,13.5])$: the situation is reversed and the profile of SIDM1b and vdSIDMb have a \emph{lower} density than the profile of CDMb ($\rho_{\text{ SIDM}}/\rho_{\text{CDM}} < 1$), but also for this group the difference between the two dark matter models is less then 20\%.
\end{itemize}

 This trend with mass mentioned above is weaker than that reported by \citet{tesi:Elliptical5}, who found steeper SIDM profiles at $M_{200}\sim10^{12.5}M_{\odot} h^{-1}$ than we do here. Given the halo-to-halo variations of the density profiles, our larger statistics allows us to better capture the properties of the entire population and avoid selection biases that can affect small samples. We also point out that the interplay between baryons could be different when using the EAGLE (as we do here) or TNG \citep[as in ][]{tesi:Elliptical5} galaxy formation models, causing systematic differences between the results. 

We find that there are no substantial differences between the two simulated SIDM models over the halo mass range considered here. This is due to the functional form of the velocity-dependent cross-section, which produces $\sigma/m \sim1\cmsg $ for $M\sim10^{13}M_{\odot} h^{-1}$ \citep[see][for more details]{robertson21}. For this reason, in the remainder on this Section we show results only for CDM and SIDM1.

We now focus on the dark matter component alone, to understand how the presence of baryons modifies the underlying distribution of dark matter: Fig.~\ref{fig:Profiles_DM} shows the dark matter density profiles in CDM (left), SIDM1 (middle) and vdSIDM (right) for the three mass bins (different colors). Solid and dashed lines stand again for the hydro and DM-only simulations and the bottom residual panels show the ratio $\rho_{\text{full}}/\rho_{\text{DM-only}}$ for each case. The presence of a central baryonic component contracts the halo in the inner region: while in CDM $\rho_{\text{full}}/\rho_{\text{only}} \sim 1.5$, the effect is much stronger in SIDM with $\rho_{\text full}/\rho_{\text only} > 4.0 $.

\begin{figure*}
    \centering
    \includegraphics[width=1.0\linewidth,height=0.6\linewidth]{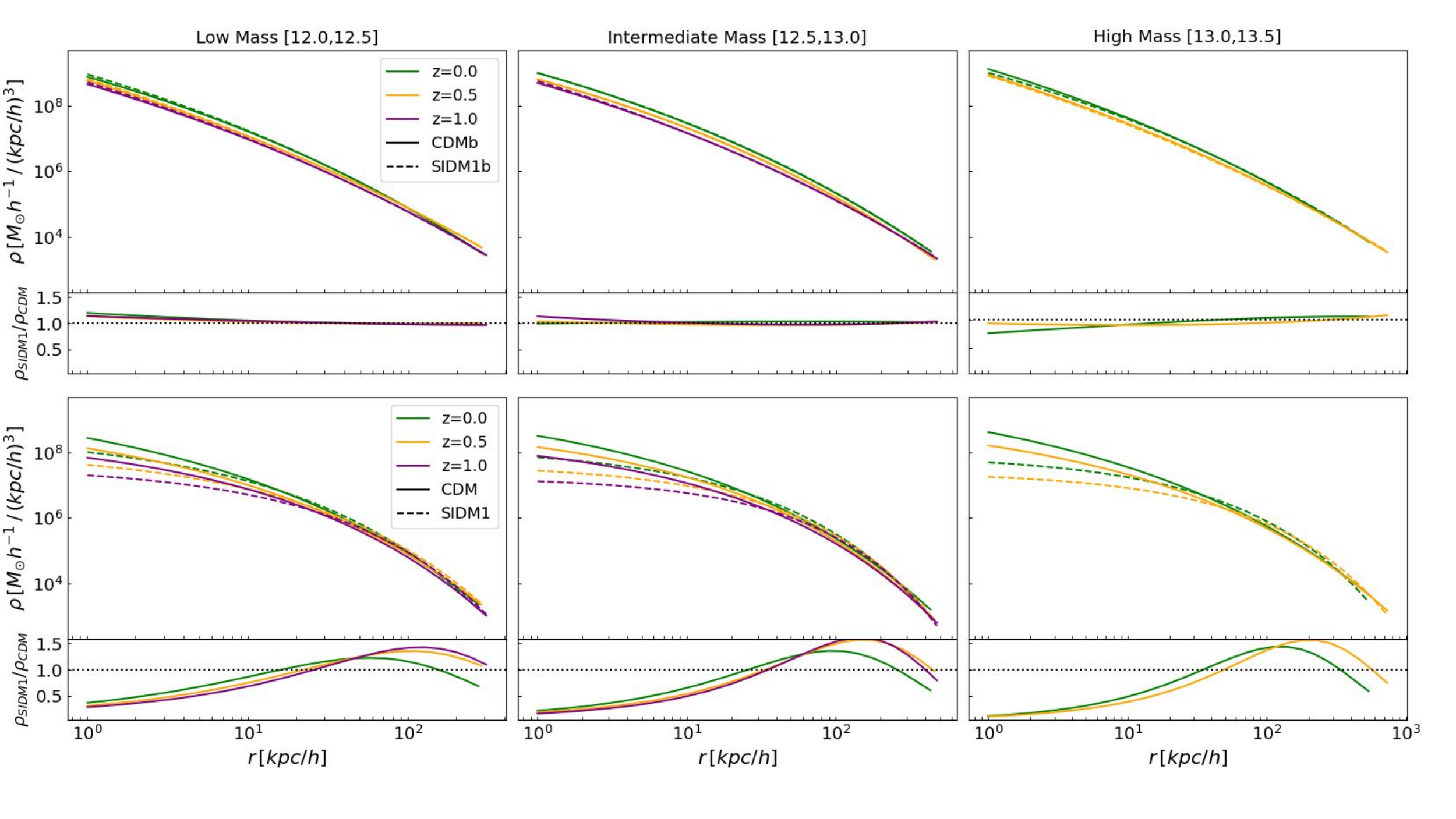}
    	\caption{Evolution with redshift of the best-fit average density profiles, considering haloes at z = 0, 0.5 and 1. Top: Full-hydro simulations. Bottom: DM-only simulations. Bottom panels: ratio between the total and the DM-only profiles in the two models: $\rho_{\text{SIDM}}/\rho_{\text{CDM}}$. For the High Mass bin, we do not show the fitted profiles at $z=1$, since we do not have enough haloes to perform a robust fit. We do not show the profiles for vdSIDM, because they are very similar to those with SIDM1.}
	\label{fig:profiles_z}

\end{figure*}

Finally, we expand the analysis to the haloes at higher redshift: $z=0.5$ and $z=1$, repeting the same procedure. Figure \ref{fig:profiles_z} shows the results from the hydro (top) and DM-only (bottom) runs. For each mass bin the different colors represent the three considered redshifts; solid lines refer to the average density profiles of the CDM model, while the dashed ones are for the SIDM1 model
The residual panels show the ratio of the SIDM1 and CDM profiles: the effects of self-interactions are almost redshift-independent, in all runs, suggesting that they depend primarily on halo mass and not on redshift.

\subsection{Concentration-mass relation}
\begin{figure*}
    \centering
    \includegraphics[width=1.0\linewidth,height=0.6\linewidth]{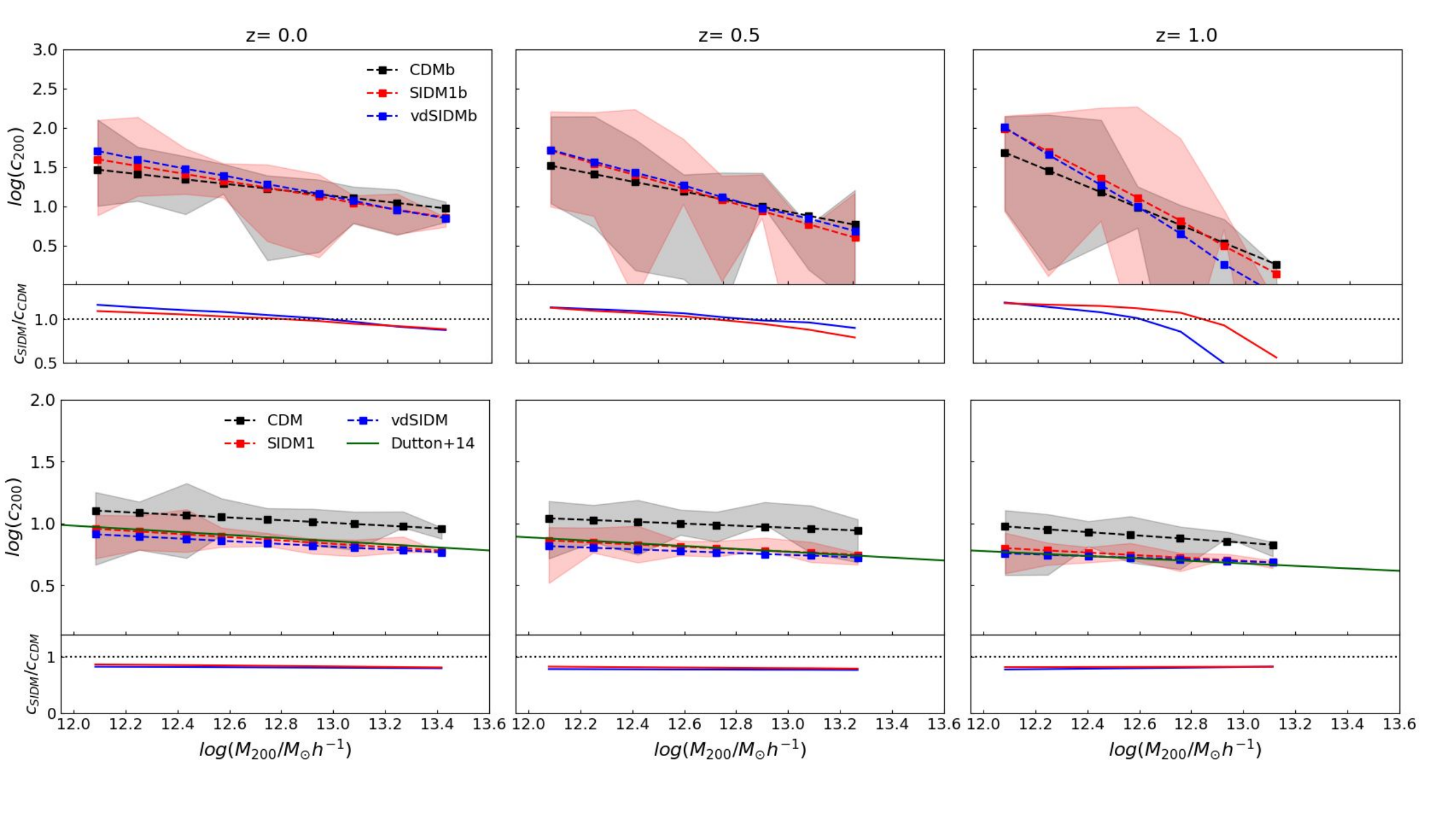}
    	\caption{Concentration-mass relation at $z=0,0.5$ and $1$. Top: Full-hydro simulations. Bottom: DM-only simulations.
    	The black, red and blue dashed lines show the best-fitting linear relation for the CDM, SIDM and vdSIDM runs, respectively. The best fit has been performed on the stacked values over different bins in mass (each containing at least two haloes). The shaded regions show the scatter in each of these bins (the vdSIDM scatter is not shown since it is similar to SIDM1).
    	The solid green lines are the best-fit relation from \citet{dutton2014cold}.
    	Bottom panels: ratio between the SIDM models and CDM model for both full-hydro and DM-only relations: $c_{\text{SIDM}}/c_{\text{CDM}}$.}
	\label{fig:Conc_mass_rel}

\end{figure*}

The profiles of individual systems can deviate from the mean profiles shown in Figure \ref{fig:densityz0} and \ref{fig:profiles_z}: the halo-to-halo variation can be characterised by means of the distribution of halo concentrations \citep{tesi:NFW,duffy2008mnras,dutton2014cold,schaller2015effect}. We now fit each halo profile with the Einasto model and use the resulting scale radius to calculate the concentration $c_{200}= r_{200}/r_{-2}$. We then define the concentration-mass relation as:
\begin{equation}
    \log (c_{200})=A-B\log (M_{200}/M_{\odot}h^{-1}).
    \label{concmass}
\end{equation}
and find the best-fit values $(A,B)$ for each considered model. These are listed in Table ~\ref{tab:5}, together with the $1\sigma$ uncertainties. We plot the best-fit c-m relations for each considered model and redshift in  Figure \ref{fig:Conc_mass_rel}.
In the DM-only runs, the three $c-M$ relations are essentially parallel to each other: the core creation induced by self-interactions reduces the halo concentration similarly for all haloes. In the hydro runs, although the scatter in the relations is higher than in the DM-only runs (shaded regions in Figure \ref{fig:Conc_mass_rel}), the SIDMb $c-M$ relations are instead steeper than than in CDMb:  haloes show a greater diversity of profiles and the $c-M$ trend is consistent with the trend in mass described in Section \ref{sec:profiles0}. Measuring the $c-M$ relation over a larger range in mass and in a larger simulation box will be essential to confirm this difference in slope.

For comparison, the concentration-mass relation of \citet{dutton2014cold}, based on DM-only simulations of a $\Lambda$CDM model with the same cosmological parameters as ours \citep{ade2014planck}, is shown as a green solid line in Figure \ref{fig:Conc_mass_rel}. Covering a wide mass range $10^{10} \leq M \leq 10^{15}\, M_{\odot}h^{-1}$ and fitting an Einasto model to estimate the concentration, they obtained $[A,B] = [2.48,0.12], \,[2.29, 0.12] \, \text{and} \, [1.97, 0.10]$, respectively for $z=0,\, z=0.5\, \text{and}\, z=1$. Although these results differ slightly from ours due to the different mass range and the different radial ranges used to perform the fits, they are consistent, within $1\,\sigma$, with the CDM concentration-mass relation that we find here.


\begin{table}
	\centering
	\caption{The linear best fit parameters of concentration-mass relations for each model at different redshift: $\log(c_{200}) = A - B\,\log(M_{200})$. See Figure \ref{fig:Conc_mass_rel}.}
	\begin{tabular}{ccccc}
		\hline
		& &  \textbf{z=0.0} &  \textbf{z=0.5} & \textbf{z= 1.0}\\
		\hline
		\\
		\textbf{CDM} & $A$ &$2.42 \pm 0.22$ & $2.04 \pm 0.54$ & $2.70 \pm 0.53$\\
		& $B$ &$0.11\pm0.02$ & $0.08\pm 0.04$ & $0.14\pm0.04$  \\
		\hline
		\\
		\textbf{CDMb} & $A$ & $5.91 \pm0.39$ & $9.27 \pm 2.18$ & $18.14 \pm 2.61$ \\
		& $B$& $ 0.37 \pm0.06$ & $0.67\pm0.16$ & $1.2\pm0.2$ \\
		\hline
		\\
		\textbf{SIDM1}  & $A$&$2.56\pm 0.11$ & $2.05 \pm 0.28$ & $2.15 \pm 0.30$ \\
		& $B$&$0.13\pm0.01$ & $0.10\pm 0.02$ & $ 0.11\pm0.02$ \\
		\hline
		\\
		\textbf{SIDM1b} & $A$ &$8.23 \pm 0.53$& $13.01 \pm 2.72$& $23.35 \pm 7.56$\\
		& $B$&$0.55\pm0.04$ & $0.94\pm0.21$ &$1.77\pm0.60$ \\
		\hline
		\\
		\textbf{vdSIDM} & $A$ &$2.24 \pm 0.09$& $1.75 \pm 0.24$& $1.61 \pm 0.30$\\
		& $B$&$0.11\pm0.01$ & $0.08\pm0.02$ &$0.07\pm0.02$ \\
		\hline
		\\
		\textbf{vdSIDMb} & $A$ &$9.34 \pm 0.99$& $12.20 \pm 1.63$& $26.59 \pm 5.97$\\
		& $B$&$0.63\pm0.08$ & $0.87\pm0.13$ &$2.04\pm0.47$ \\
		\hline

	\end{tabular}
	\label{tab:5}\centering
\end{table}

\section{Dark matter fractions} \label{sec:fractions}

A direct consequence of the different composition of the central part of haloes is a systematic difference in the dark matter fraction, that can be lower in self-interacting models with respect to CDM. In this Section, we attempt a comparison of the simulated dark matter fractions with observational results and we discuss the possibility of using observed fractions to discriminate between the dark matter models considered here.

Previous numerical works \citep[e.g.][]{lovell18} have calculated the dark matter fraction, or conversely the baryon fraction, at the center of simulated haloes and compared it to observational results from lensing \citep{auger10b,barnabe11,sonnenfeld13}, Jeans modelling \citep{tortora12} and survey data \citep{cappellari13}, among others. Observed dark matter fractions span a wide range of values, depending on the method and model used to derive them, and they can appear inconsistent with predictions from numerical simulations and with each other. This could be due to a number of reasons, including the details of the IMF model, and the techniques and apertures used for the measurements: it can thus be tricky to reproduce exactly the same measurement procedures in simulations and observations. However, another possibility is that dark matter models different from CDM could be a better match to observational measurements. Here we test this hypothesis for the case of SIDM1: even in the presence of a bias between observed and simulated fractions, we want to test if simulated CDMb and SIDM1b values follow a different distribution. In this respect, it is less important to demonstrate that simulated fractions exactly reproduce the observed distribution, and instead we search for systematic differences between CDM and SIDM1 simulated values.

\begin{figure*}
    \centering
    \includegraphics[width=\linewidth]{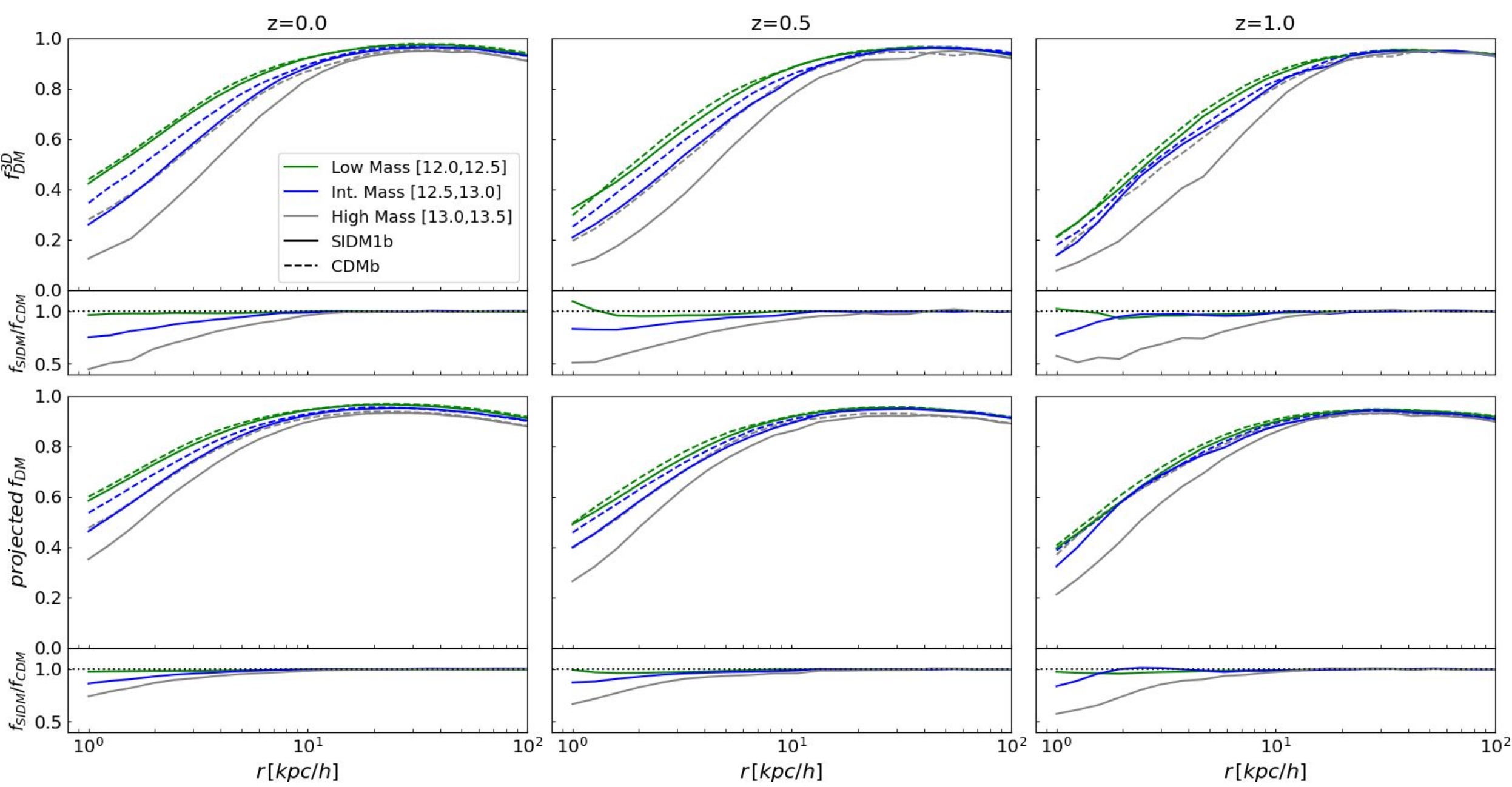}
    \caption{Fraction of enclosed dark-matter mass, $f_{\text{DM}}$, as a function of radius, for each mass bin (different colours) and for the CDMb (solid lines) and SIDM1b (dashed lines) runs. Top: Intrinsic 3D dark matter fraction. Bottom: projected DM fraction, calculated by averaging over three viewing angles for each halo.}
    \label{fig:dmfrac1}
\end{figure*}

\begin{figure}
    \centering
    \includegraphics[width=\linewidth]{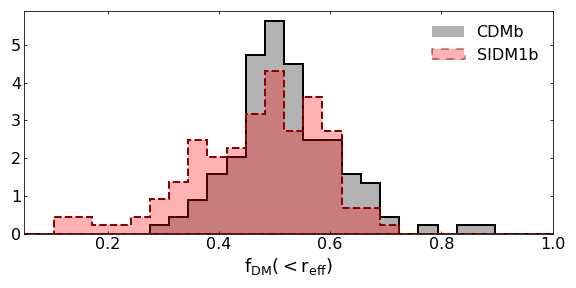}
    \includegraphics[width=\linewidth]{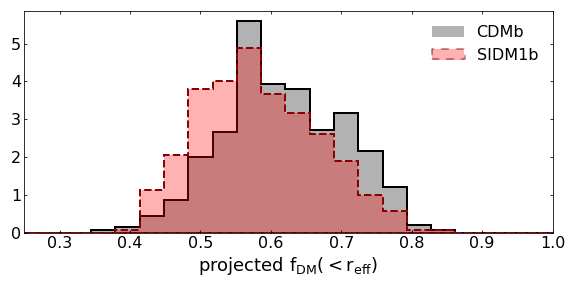}
    \caption{Normalised distribution of simulated dark matter fractions in the CDMb and SIDM1b models. In the top panels, we use the $z=0$ 3D dark matter fraction $f_{\text{DM}}$  within the effective radius $r_{\text{eff}}$, while in the bottom panel we show instead the \emph{projected} dark matter fraction at $z=0.5$. }
    \label{fig:dmfrac3}
\end{figure}

We calculate the intrinsic 3D and projected dark matter fractions $f_{\text{DM}}$ as a function of distance from the halo center for the two hydro runs (i.e. CDMb and SIDM1b) and the three considered redshifts. In projection, we use three viewing angles for each halo to increase our statistics. The mean $f_{\text{DM}}$ profiles are shown in Figure \ref{fig:dmfrac1}.
While for the Low Mass bin, CDMb and SIDM1b predictions are in practice identical, the highest-mass haloes show central fractions 50 per cent lower in SIDM1b than CDMb, consistent with what we found in Section \ref{sec:profiles}. However, the difference is significant only in the inner $\sim$10 kpc and less pronounced in projection (bottom panels). 
To quantify the effect on a scale that is common observational studies, in Figure \ref{fig:dmfrac3} we plot the normalised distribution of 3D (top) and projected (bottom) dark matter fraction calculated within the effective radius of the central galaxies $r_{\text{eff}}$, calculated as the radius that contains half of the stellar mass. In the simulations, this is simply done from the star particles, while observations rely on the galaxy light to measure the effective radius. In Figure \ref{fig:dmfrac3} we notice a shift between the two distributions, with self-interactions causing a more important tail at low fractions, both in 3D and in projection. 

\begin{figure*}
    \centering

    \includegraphics[width=0.75\linewidth]{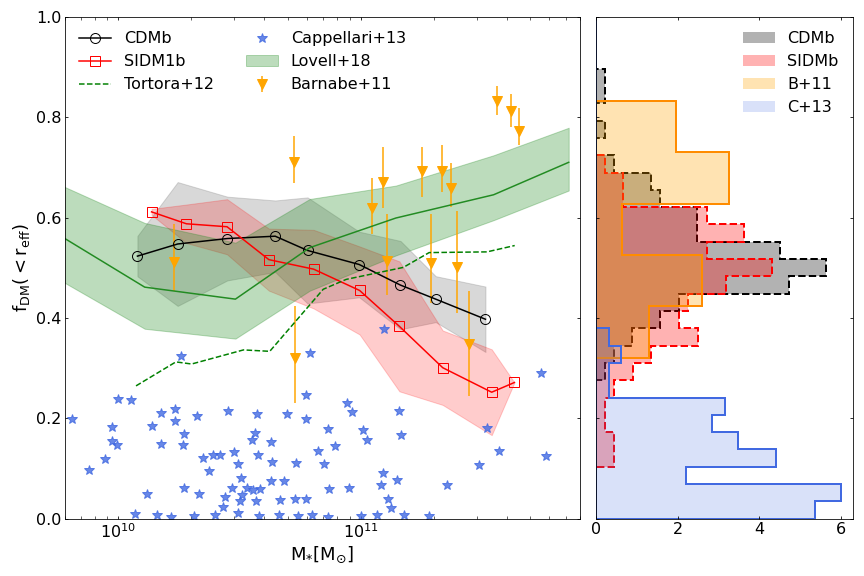}
    \includegraphics[width=0.76\linewidth]{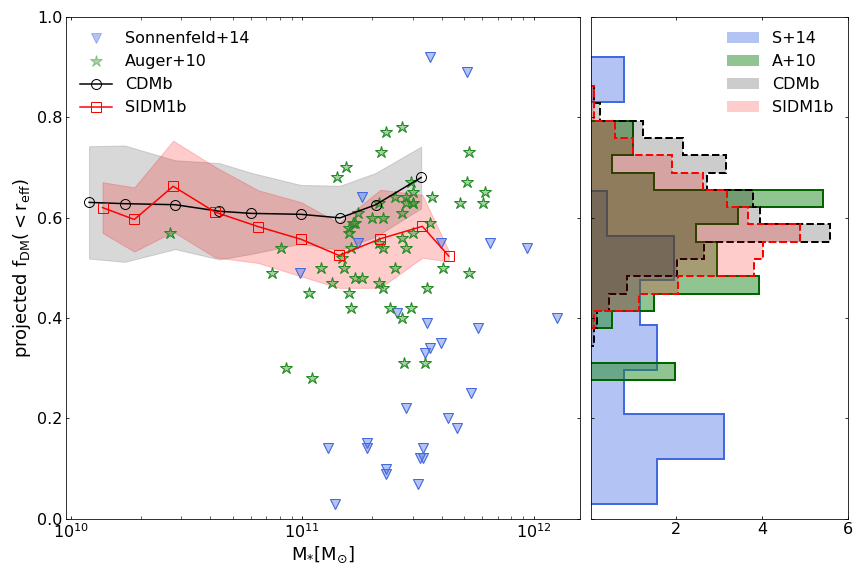}
    \caption{Comparison to observational results at $z=0$ (top) and $z=0.5$ (bottom). In the top-left panel, we plot the $z=0$ 3D dark matter fraction $f_{\text{DM}}$  within the effective radius $r_{\text{eff}}$, measured in the two hydro runs as a function of the galaxy stellar mass; the CDMb and SIDM1b distributions are represented respectively by the black circles and red squares, together with the $1\sigma$ (shaded) region. We compare them with results from previous works that use simulations \citep[][the TNG runs]{lovell18} or observations of Early-Type galaxies with: gravitational lensing \citep[][yellow triangles]{barnabe11}, dynamical (Jeans) modelling \citep[][dashed line]{tortora12} and survey data \citep[][blue stars]{cappellari13}. In the bottom, we show instead the \emph{projected} dark matter fraction at $z=0.5$, compared to values derived with gravitational lensing for the SLACS \citep[][green stars]{auger10b} and SL2S \citep[][blue triangles]{sonnenfeld13} samples. For the simulated data we consider three projections per halo. In the right smaller panels, we compare the normalised distributions of simulated and observed fractions to each other.}
    \label{fig:dmfrac2}
\end{figure*}

In Figure \ref{fig:dmfrac2}, we then compare the same dark matter fractions to previous results, by plotting them as a function of the stellar mass of the central galaxy. In the top panel, we plot the 3D fractions calculated at $z=0$ within the effective radius $r_{\text{eff}}$ as a function of stellar mass from the CDMb (black dots) and SIDM1b (red squares) runs. The green solid line (and band) shows the results obtained for bulge-dominated galaxies by \citet{lovell18} with the IllustrisTNG runs (here we do not reproduce the same selection based on galaxy type). We then plot observational values obtained from $(i)$ the analysis of the SLACS sample of gravitational lenses at $z<0.3$ \citep[yellow triangles]{barnabe11}, $(ii)$ dynamical modelling \citep[green dashed line]{tortora12} and $(iii)$ elliptical $z=0$ galaxies from the ATLAS$^{3D}$ survey \citep[blue stars]{cappellari13}. In the bottom panel we show instead the \emph{projected} dark matter fractions at $z=0.5$, i.e. calculated inside a cylinder of aperture $r_{\text{eff}}$; in this case, for each simulated halo we use three projections that we treat as independent measurements. We compare these values to equivalent measurements in two samples of gravitational lenses: the SLACS \citep[green stars]{auger10b} and the SL2S \citep[blue triangles]{sonnenfeld13} lenses. 

From the mean fractions (black and red lines in both main panels), we notice a different trend of the $f_{\text{DM}}-M_{*}$ relation: self-interactions produce lower dark matter fractions at the high mass end, and the trend is reversed at low masses. This is consistent with the mass-dependent differences in the density profiles and in the dark matter fraction profiles, seen in Section \ref{sec:profiles}. In projection (bottom panel), the difference between the two is smaller and the mean $f_{\text{DM}}$ is somewhat flatter. 

Finally, the two right panels of Figure \ref{fig:dmfrac2} show the normalised distribution of the dark-matter fractions, comparing simulations and observations in order to point out sample differences.
While the lens sample from \citet{auger10b} is consistent with the simulated fractions, \citet{sonnenfeld13} and - especially - \citet{cappellari13} derived much lower dark matter fractions. It is difficult to reproduce exactly the distribution measured in observations: possible sources of bias include the sample selection, the difference between the true value of $r_\text{eff}$ known from simulations and the equivalent measurement from the data, or the spread in redshift of the observed data points. Nevertheless,  it is clear that the scatter of observed data points is larger than that produced by the different dark matter models: a signature of self-interactions on dark matter fractions exists, but a more detailed evaluation of the observational biases would be required to reach a definitive conclusion.

With our measurements, we show that self-interacting models can produce lower dark matter fractions at the high mass end, widening the predicted distribution of $f_{\text{DM}}$. However, we do not predict values as low as some of the observations and thus the difference has to be searched elsewhere, i.e. in biases between predicted and observed quantities or in more extreme alternative dark matter models. If these were resolved and measurements were more precise, the offset between the $f_{\text{DM}}$ distributions (Figure \ref{fig:dmfrac3}) could be used to discriminate between CDM and alternative SIDM models.

\section{Strong lensing effect} \label{sec:lensing}

We now investigate the effect of SIDM on the lensing properties of simulated haloes. Gravitational lensing is one of the most accurate techniques to measure the total mass of galaxies and clusters, as well as their radial distribution. The effect of warm or self-interacting dark matter on the mass distribution and thus on the lensing signal of haloes and subhaloes has been studied in previous works \citep{robertson18b,despali20,gilman20,hsueh20}. However, the number of haloes available in the runs used here gives us a chance to derive more precise predictions for the population of massive galaxies.

Starting from the particle distribution, we create 2D maps of the projected mass distribution using the library Py-SPHViewer \citep{SPHviewer}, as done by \cite{ME2020} \citep[see also ][]{ME2022}. From these, we calculate the lensing convergence $\kappa$ - the Laplacian of the lensing potential - with the {\sc PyLensLib} library \citep{Meneghetti2021}. The value of the lensing convergence determines by how much the background sources appear magnified on the lens plane.  In practice, the convergence is defined as a dimensionless surface density and so effectively corresponds to a scaled projected mass density, characterising the lens system. It can be written as
\begin{equation}
\kappa(x)= \frac{\Sigma(x)}{\Sigma_\rmn{crit}}, \qquad \rmn{with} \qquad \Sigma_\rmn{crit}= \frac{c^{2}}{4\pi G} \frac{D_\rmn{S}}{D_\rmn{LS}D_\rmn{L}}, \label{eq:lens}
\end{equation}
where $\Sigma_\rmn{crit}$ is the critical surface density and $D_{L}$, $D_{S}$ and $D_{LS}$ stand respectively for the angular diameter distance to the lens, to the source, and between the lens and the source. 
Finally, as an estimate of the lensing power of each halo, we calculate the size of the largest critical curve of each system, i.e. the region of the plane where the magnification is formally infinite, corresponding to the region where the lensed images form. A massive galaxy can have more than one primary critical line if the system is composed of multiple mass components.    Our criterion to identify the primary critical lines is based on the size of the effective Einstein radius: given a critical line enclosing the area $A_c$, the Einstein radius can be calculated as $\theta_{\text{E}} = \sqrt{A_c/\pi}$.

\begin{figure*}
    \centering
    \includegraphics[width=\linewidth]{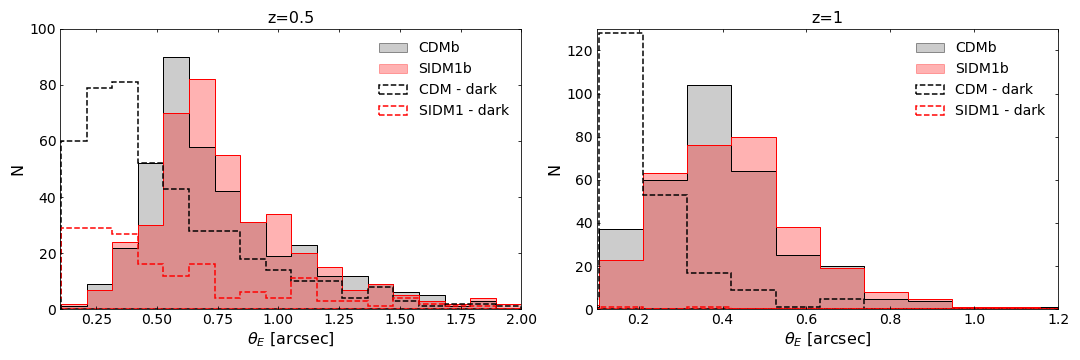}
    \caption{Sizes of the Einstein radii $\theta_{\text{E}}$ from the lensing convergence maps. We plot the distribution of the inferred sizes for the entire sample of haloes in the different runs. We consider three projections per halo and two redshift configurations that corresponds to observed typical cases: lens systems at redshift $z=0.5$ and sources at $z=2.5$ (shown on the left) and lenses at redshift $z=1$, with sources at $z=3$ (on the right). The CDM and SIDM1 results are shown in black and red, respectively; empty dashed histograms stand for the results of the dark-matter-only runs, while filled histograms show the hydro case. At $z_{L}=1$, the SIDM1 dark run produces critical curves only in a handful of cases and the $\theta{\text{E}}$ distribution is thus barely visible in the plot. }
    \label{fig:conv_sizes}
\end{figure*}

We repeat this process for all runs and use three projections per halo (the same used to calculate the projected dark matter fractions), in order to increase the sample size. The resulting distributions of Einstein radii $\theta_{\text{E}}$ are shown in Figure \ref{fig:conv_sizes}: in the two panels, we show the results of two different choices of lens and source redshift that reproduce common observed galaxy-galaxy lensing configurations. The haloes used to create the lensing convergence have been selected from the snapshot that correspond to the lens redshift, while the source redshift is used in the calculation of $D_{S}$ and $D_{LS}$ and thus influences the convergence. 

From the distributions, it is clear how the similarity between the halo profiles in the hydro runs propagates into similarities in their lensing properties, in comparison to the dark-matter-only runs (empty dashed histograms). Counter-intuitively, a small shift between the two distributions can be seen (especially in the right hand panel), producing a lack of small Einstein radii in SIDM1. This fact is a result of the different slope of the concentration-mass relations: the least massive haloes in our sample are more concentrated in SIDM1b than CDMb and thus produce a stronger lensing effect. However, this does not influence the high-$\theta_{\text{E}}$ end of the distribution, dominated by massive galaxies.

It is worth pointing out that the SIDM1 dark-matter-only run not only predicts smaller-separation lensed images, but quite often does not produce any strong lensing at all. In the left panel of Figure \ref{fig:conv_sizes}, the total number of critical lines produced in SIDM1 is significantly lower than the other three considered models (with a comparable total number of objects) and the configuration chosen in the right panel does not produce any critical curves. These results are similar to the findings from \citet{robertson18b} on cluster scales and demonstrate that a large sample size is essential to obtain reliable predictions.

\begin{figure}
    \centering
    \includegraphics[width=\linewidth]{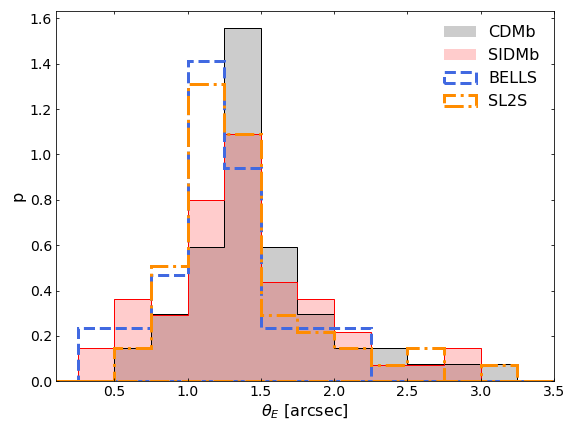}
    \caption{Comparison between the lensing properties of simulated haloes and real lens samples. Similarly to Figure \ref{fig:conv_sizes}, the black and red filled histograms show the distribution of $\theta_{\text{E}}$ calculated from the CDMb and SIDM1b runs, respectively. However, here we use only the systems with a mass $M_{\text{200}}>10^{13}M_{\odot}h^{-1}$ and we show only the results for $z=0.5$ (with $z=2.5$ as the source redshift), for a more precise comparison with the two observational samples. These are the BELLS-Gallery \citep{ritondale19a} and SL2S \citep{sonnenfeld13} lens samples, that include massive elliptical galaxies at $z_{l}\sim0.5$ and $0.5\leq z_{l} \leq 0.8$, with sources at $z_{S}\sim2.5$. We compare the normalised distribution of Einstein radii derived in previous works to the sizes calculated from our simulations, finding a good agreement.}
    \label{fig:conv_sizes2}
\end{figure}

Finally, we compare the distribution of $\theta_{\text{E}}$ that we obtain from the hydro runs to observed values. In particular, we consider two lens samples: the BELLS-Gallery \citep{ritondale19a} and SL2S \citep{sonnenfeld13} lenses, that include massive elliptical galaxies at $z_{l}\sim0.5$ and $0.5\leq z_{l} \leq 0.8$, with sources at $z_{S}\sim2.5$; previous works provide us with the inferred values of the Einstein radius for each system. We thus select haloes at $z=0.5$ and with total masses $M_{\text{200}}>10^{13}M_{\odot}$, corresponding to the mass range of ETGs. Figure \ref{fig:conv_sizes2} shows the normalised distribution of $\theta_{\text{E}}$ for the two simulations and the two observed samples. We find a good agreement between the distributions, which is an encouraging indication of the fact that simulated systems are able to reproduce observed quantities. We point out again that, in this work, we do not explicitly select haloes on the basis of galaxy morphology, while observed lenses are typically elliptical galaxies. However, the cut in halo mass allows us to select the most massive haloes, which most probably host massive ellipticals.

From Figure \ref{fig:conv_sizes2}, we conclude that we cannot distinguish between the cold and self-interacting hydro models by looking at the simulated sizes of the lensed images. This is somewhat in contrast with the results from \citet{tesi:Elliptical5}, who found a different distribution in the two models. While we cannot exclude that part of the difference is due to the different code and hydro model used for the simulations (Gadget with EAGLE model  vs Arepo with TNG model), the main difference between the two is the considered sample: \citet{tesi:Elliptical5} used only nine galaxies, while here we have more than 100 systems per model and redshift. Moreover, the density profiles in \citet{tesi:Elliptical5} showed a clear trend in SIDM with respect to CDM - i.e. they were either cored or more cuspy than their CDM counter parts - while here we observe a larger range thanks to the improved statistics, more representative of the entire population.

\section{Conclusions} \label{sec:conclusions}

SIDM has become an attractive alternative to CDM, due to its ability to produce a wider range of configurations that could solve the small-scale CDM problems, leaving the properties on large scale unchanged. However, these predictions have so far been based on DM-only simulations, while it is essential to understand how the interplay between self-interaction and baryonic effects can further modify the properties of haloes. For this purpose, we use EAGLE cosmological simulations, that simulate the same volume of 50 Mpc$^3$, in CDM and the two SIDM models: one with constant cross section ($\sigma/m = 1 \cmsg$) and one with a velocity-dependent cross section. For each model, a DM-only and a full-hydrodynamic version is available, allowing us to study the effect of baryons and alternative dark matter models at the same time.

We select haloes in the mass range between $M_{\odot} h^{-1}10^{12}\leq M_{200}\leq 10^{13.5} M_{\odot} h^{-1}$ at $z = (0,0.5,1)$. 
We split our sample into mass bins in order to compare and average the features of haloes of comparable mass. To have a sufficient number of haloes for each bin, we divide our sample into 3 bins: $12.0 < \log(M_{200}/M_{\odot}h^{-1}) < 12.5$ (Low Mass), $12.5 < \log(M_{200}/M_{\odot}h^{-1}) < 13.0$ (Int. Mass) and $13.0 < \log(M_{200}/M_{\odot}h^{-1})  < 13.5$ (High Mass) - given the volume of the simulations, we do not find higher masses.

We analyse the halo density profiles and concentrations, finding best-fit parameters for the Einasto profiles and concentration-mass relation that describe our data. We proceed by measuring the dark matter fractions in the hydro runs and the predicted sizes of lensed images, if our haloes generated galaxy-galaxy lensing events. Below we summarize our main results.
\begin{itemize}
    \item In agreement with previous work, we find that in DM-only runs the spherically averaged density profiles of SIDM (SIDM1 and vdSIDM) haloes produce a central core with density lower by 50\% (or more) compared to the CDM case. We find that the inclusion of baryons reduces these differences between the density profiles in different dark matter models (to less then 20\%). Despite this, we find a weak trend with mass of the average final properties of SIDM density profiles in full-hydrodynamic runs: the most massive systems show cored profiles, while less massive ones have cuspier profiles. We show that these properties are redshift-independent in the range $0\leq z\leq1$.
    \item We fit the density profiles with the Einasto model and we repeat the fit twice, i.e. leaving the shape parameter $\alpha$ free to vary or fixing it to $\alpha=0.16$. We find that the profiles with fixed $\alpha$ well describe CDM haloes and all full-hydrodynamic runs, but fail to reproduce the central core for the dark-matter-only SIDM runs (higher values are needed).
    \item The concentrations of CDM and SIDM haloes, at fixed $z$, decrease monotonically with increasing halo mass. In the DM-only runs, we find that the $c-M$ relations have a similar slope in all models, but SIDM concentrations are lower than CDM ones: the core creation induced by self-interactions reduces the halo concentration similarly for all haloes. In the hydro runs, the SIDM relations are instead steeper than in CDM, due to the greater diversity of profiles and the dependence of mass found earlier. These differences translate into a small shift between the Einstein radius distributions: the least massive haloes in our sample are more concentrated in SIDM1b than CDMb and thus produce a stronger lensing effect.
    \item We calculate the dark matter fractions within the effective radius of the central galaxies $r_{\text{eff}}$ and compare them to observational results. We find that at the high mass end, SIDM models can generate lower dark matter fractions. However, we do not predict values as low as some of the observations and thus the difference between the distributions of simulated and observed DM fractions has to be searched for elsewhere, i.e. in biases between predicted and observed quantities or in more extreme alternative dark matter models. Methods able to distinguish between the dark matter and baryonic components in galaxies and clusters, performing a so-called ``mass-decomposition'', constitute one of the most promising paths to discriminate between CDM and SIDM.
    \item We compare the distribution of Einstein radii that we obtain from the hydro runs to observed values. We find a good agreement between the distributions, which is an encouraging indication of the fact that simulated systems are able to reproduce observed quantities. However, we conclude that we cannot distinguish between the CDM and SIDM hydro models by looking at the simulated sizes of the lensed images.

\end{itemize}

In this work, we have analysed for the first time the properties of massive galaxies in a cosmological box that includes both self-interacting dark matter and baryons. In agreement with previous works, we find that the halo properties predicted in SIDM are deeply influenced by  the inclusion of baryons. Recently, \citet{eckert2022constraints} performed a similar analysis on the BAHAMAS-SIDM cluster-scale simulations \citep{mccarthy2016bahamas, robertson18b}. Through a comparison with the density profiles of observed galaxy clusters, they were able to put an upper limit on the self-interaction cross section of $\sigma/m <0.19\cmsg$. Here, on the other hand, we cannot exclude the higher value of $\sigma/m\sim1\cmsg$. In particular, we find that the SIDM models considered here cannot be ruled out at the scale of massive galaxies, contrary to prevous claims. \citet{eckert2022constraints} also measured the Einasto shape parameter $\alpha$ at cluster scales and found that (in hydro simulations) it differs in different DM scenarios, and can be used to set constraints, while in our case the distribution of $\alpha$ is very similar with CDM and SIDM (see Figure \ref{fig:alpha}). This supports the hypothesis that a self-interacting model with a stronger velocity dependence could explain well the data on both scales, as recently suggested by other works \citep{correa21,adhikari22}.  It is fundamental to move forward in the field and investigate the interplay between baryons and alternative dark matter, as it has successfully been done in the past for the CDM model. 

\section*{Acknowledgements}
This work used the DiRAC@Durham facility managed by the Institute for Computational Cosmology on behalf of the STFC DiRAC HPC Facility (www.dirac.ac.uk). The equipment was funded by BEIS capital funding via STFC capital grants ST/K00042X/1, ST/P002293/1 and ST/R002371/1, Durham University and STFC operations grant ST/R000832/1. DiRAC is part of the National e-Infrastructure.
LM acknowledges the support from the grant ASI n.2018-23-HH.0 and from the grant PRIN-MIUR 2017 WSCC32.

\section*{Data Availability}

The catalogues and data products used for this paper can be made available upon reasonable request to the corresponding author.



\bibliographystyle{mnras}
\bibliography{example} 





\bsp	
\label{lastpage}
\end{document}